# Controlled variation of the refractive index in ion-damaged diamond

P. Olivero[1,2], S. Calusi[2,3], L. Giuntini[3], S. Lagomarsino[4],

A. Lo Giudice[1,2], M. Massi[3], S. Sciortino[4], M. Vannoni[5], E. Vittone[1,2]

[1] Experimental Physics Department and "Nanostructured Interfaces and Surfaces" Centre of Excellence, University of Torino, via P. Giuria 1, 10125 Torino, Italy

[2] INFN Sezione di Torino, via P. Giuria 1, 10125 Torino, Italy

[3] Physics Department of University and INFN, Firenze, via Sansone 1, 50019 Sesto Fiorentino (Firenze), Italy

[4] Energetics Department of University and INFN, Firenze, via Sansone 1, 50019 Sesto Fiorentino (Firenze) Italy

[5] CNR, Istituto Nazionale di Ottica Applicata (INOA), Largo E. Fermi 6, 50125 Arcetri, Firenze Italy

**Abstract**

A fine control of the variation of the refractive index as a function of structural damage is essential in the fabrication of diamond-based optical and photonic devices. We report here about the variation of the real part of the refractive index at $\lambda$=632.8 nm in high quality single crystal diamond damaged with 2 and 3 MeV protons at low-medium fluences ($10^{13}$ - $10^{17}$ ions cm$^{-2}$). After implanting the samples in 125×125 μm$^2$ areas with a raster scanning ion microbeam, the variation of optical thickness of the implanted regions was measured with laser interferometric microscopy. The results were analyzed with a model based on the specific damage profile. The technique allows the direct fabrication of optical structures in bulk diamond based on the localized variation of the refractive index, which will be explored in future works.

**Introduction**

Since many years diamond has been considered the ultimate material in many optical applications. Diamond optical devices are ideal in extreme physical conditions where low optical absorption is required together with efficient heat dissipation and structural resistance (i.e. laser fan-out and diffractive elements, RF power optics, high-vacuum windows, optical micro-electromechanical systems, etc) [1-8]. More recently, diamond attracted wide interest as a prominent candidate for the implementation of quantum information processing schemes with the employment of defect-based single color centers as single photon emitters or quantum bit storage elements [9-11]. In 2009 it was proposed that the local alteration of the refractive index can be employed in the fabrication of diamond-based photonic devices, such as photonic crystals and high-Q microcavities [12].

The variation of refractive index of single crystal diamond as a function of structural defect density has been investigated in a surprisingly limited number of works [13-16] and no systematic data have been reported on the variation of the refractive index in diamond in the low-damage-density regime. Our paper represents a systematic study of the variation of the real part of the refractive index of single diamond at $\lambda$=632.8 nm as a function of MeV-ion-induced structural damage. This specific wavelength is very close to the NV⁻ emission, which is of extreme interest in the emerging field of diamond-based quantum optics and photonics [9-11].

**Experimental**

This study was carried out on two 3.0×3.0×0.5 mm$^3$ samples of type IIa single-crystal diamonds grown with Chemical Vapour Deposition (CVD) technique by ElementSix [17]. The crystals consist of a single {100} growth sector, with concentrations of nitrogen and boron impurities below 0.1 ppm and 0.05 ppm, respectively. The crystals are cut along the <100> axes and the two opposite faces of the samples are optically polished.

The samples were implanted at the external scanning microbeam facility of the LABEC laboratory in Firenze [18]. The beam was focused on the polished side of the samples to a spot of ~10 μm and ~20 μm, for 3 and 2 MeV protons, respectively.

The ion current varied between 0.2 nA and 1.5 nA. For each implantation, the ion beam was raster scanned over a square area of ~125×125 μm$^2$, in order to deliver a homogeneous fluence in the central region; the samples were implanted at fluences ranging from ~10$^{13}$ cm$^{-2}$ to ~10$^{17}$ cm$^{-2}$. The beam charge was measured in real time by monitoring the X-ray yield from the beam exit window [18], resulting in ~3% accuracy in the determination of the implanted charge. The actual size of the implanted area was subsequently evaluated for each implantation as the area in which the measured optical path difference (OPD) lies above the half-maximum value. The sizes dispersion (about 1.5%) was assumed to be equal to the uncertainty in the determination of the implanted area. Thus, the resulting error in the determination of fluence does not exceed 5%.

After implantation, the phase shift of a laser beam crossing the sample thickness perpendicularly to the implanted surface was determined with an interferometric technique, as shown schematically in Fig. 1. A laser interferometric microscope (Maxim 3D, Zygo Corporation) was employed, whose configuration can be briefly summarized as follows. After crossing a 20× microFizeau objective, an expanded $\lambda$=632.8 nm He-Ne laser beam is splitted into two distinct paths. The test beam crosses the sample along the thickness direction and after recombining with the reference beam produces an interference pattern which is recorded with a CCD camera; by using a phase shift method, the relative phase of the test beam is retrieved [19]. In the sample surface plane (i.e. x, y axes), the field of view was 349×317 $\mu m^2$ with an optical resolution of 1.68 $\mu$m, and the interferometric resolution in the thickness direction (i.e. z axis) was 0.63 nm. The method allows the acquisition of a map of the OPD from the region of interest (i.e. the implanted area) with micrometer lateral resolution and sub-nanometer accuracy in the optical thickness direction. It is worth stressing that the OPD signal is relevant to the whole sample thickness crossed by the test beam, as it is not affected by optical absorption process. A typical map is reported in Fig. 2(a): a significant change in OPD (~300 nm) is measured in correspondence of the implanted area (the instrumental noise is of the same order of the surfaces roughness, i.e. ~2-5 nm). The change in OPD is to be attributed to the change in optical thickness with respect to the unimplanted surrounding area. MeV ion implantation is associated with surface swelling, due to the significant variation of density in the ion-damaged material [20, 21]. As shown in Fig. 2(b), the swelling effect measured with a white light interferometric profilometer (Zygo Newview) amounts to no more than 10% of the measured OPD and its contribution has been taken into account by subtracting to the measured OPD the swelling height times the difference between the refractive index of undamaged diamond and that of air.

**Results and discussion**

The trend of the OPD profiles, corrected for the swelling, reveals a systematic dependence from the implantation fluence for 2 and 3 MeV protons, as shown in Figs. 3(a) and 3(b) respectively. In order to obtain the dependence of the refractive index from the radiation-induced vacancy density, the experimental data were analyzed as follows. In our model we assumed, at each depth, a linear dependence of the vacancy density from the implantation fluence, as resulting from the SRIM Monte Carlo simulation code [22], where the atomic displacement energy was set to 50 eV [23]. The density of vacancies per unit length induced by a single ion with energy $E$ is defined as $p_E(z)$. Such a profile is strongly non-uniform for MeV ions, as shown in Fig. 4. Non-linear processes, such as self-annealing, ballistic annealing and defect interaction were not taken into account; it has been shown that at fluences that do not exceed the graphitization limit, such an approach provides an adequate description of the ion-induced damage process in diamond in many respects [24]. Under this assumption, the vacancy density induced at depth $z$ by an implantation of fluence $\phi$ is given by the product $v(z) = \phi \times p_E(z)$. Under the assumption that the refractive index $n(z)$ depends from the vacancy density $v(z)$ at the same depth $z$, let us consider the polynomial expansion to the first order of the refractive index as a function of the vacancy density at a defined depth $z$:

$$n(z) = n_0 + \sum_{m=1}^{\infty} c_m \cdot v(z)^m \cong n_0 + c_1 \cdot v(z) \tag{1}$$

Thus, the OPD accumulated when crossing the implanted material ($0 \leq z \leq d$) is given by:

$$OPD^{(E)}(\phi) = \int_0^d [n(z) - n_0] dz \cong c_1 \cdot I_1^{(E)} \cdot \phi \qquad I_1^{(E)} = \int_0^d p_E(z) dz \tag{2}$$

In equation (2), the optical path difference ($OPD$) and the implantation fluence ($\phi$) represent the experimental data and the $I_1^{(E)}$ coefficients are evaluated by integrating the profiles generated by SRIM simulation for each ion energy. Therefore, in this linear approximation, the ($OPD$ vs $\phi$) plots reported in Fig. 3 for the two ion energies can be fitted with linear functions, whose coefficients allow the evaluation of the $\{c_1\}$ parameters by taking into proper account the different vacancy generation profiles via the $I_1^{(E)}$ coefficients. Limiting the expansion of expression (1) to the first-degree is well justified for the data relevant to 2 and 3 MeV proton implantations, as it is apparent from the fitting curves reported in Fig. 3 (linear correlation coefficients exceeding 99%). It is worth stressing that, although the two ($OPD$ vs $\phi$) curves have different slopes, as expected for different damage profiles, the resulting $c_1$ parameters are consistent within the uncertainties: $c_1 = (4.30 \pm 0.06) \cdot 10^{-23}$ cm$^3$ for 2 MeV implantations, and $c_1 = (4.26 \pm 0.12) \cdot 10^{-23}$ cm$^3$ for 3 MeV implantations. This result validates the assumption that, within the investigated range of fluences, the variation of refractive index depends linearly from the structural damage as evaluated with the vacancy density predicted by the SRIM code.

From the weighted mean of the previous $c_1$ values, the dependence of the refractive index at λ=632 nm from the vacancy density can be obtained as follows:

$$\Delta n = (4.3 \pm 0.3) \cdot 10^{-23} \times v \left[ cm^{-3} \right] \tag{3}$$

where the overall uncertainty has been obtained by propagating the uncertainties from the fit procedure (1%), the implanted charge calibration (3%), the determination of the implanted area (1.5%) and the estimation of the vacancy/ion generation from SRIM (5%).

In our study, the maximum variation of the refractive index amounts to ~0.1, corresponding to ~4% of the absolute value at λ=632 nm (*n*=2.41). It is worth stressing that our observation of the increase of the refractive index of diamond at low damage levels is consistent with what reported in [13].

**Conclusions**

In the present work we report about the controlled variation of the refractive index of diamond damaged by MeV protons at medium-low fluences. If it is assumed that no specific structural defects are responsible for the variation of refractive index, and thus that generic structural damage (quantified in a density of vacancies in SRIM simulations) can be adopted as a general effective parameter, then from the consistency between data relevant to 2 and 3 MeV protons, the formula (3) can be applied for any damage profile generated in diamond. Alternatively, if it is assumed that specific damage process are involved in the variation of the refractive index, at the very least the formula is expected to hold true for implantation of light MeV ions, provided that the damage density does not exceed ~3·$10^{21}$ vacancies cm$^{-3}$, corresponding to the maximum density of vacancies generated in our experiment, as evaluated by SRIM code. We envisage that the functional dependence that we identified can be directly applied to the engineering and fabrication of a vast range of optical structures based on refractive index contrast in single crystal diamond with MeV ion implantation, which will be investigated in future works.

**Acknowledgments**

This work is supported by the "Accademia Nazionale dei Lincei – Compagnia di San Paolo" Nanotechnology grant and by "DANTE" and "FARE" experiments of "Istituto Nazionale di Fisica Nucleare" (INFN), which are gratefully acknowledged. The authors thank F. Gucci, A. Catelani and M.Falorsi for technical advices and help.

**Figure captions**

Fig. 1 (Color online) Schematics and principle of operation of the laser beam interferometric microscope. An expanded laser beam crosses a first beam splitter, then after the microFizeau objective it is splitted into two distinct paths when crossing a "reference surface". The first path ("reference beam") is re-directed towards the detector, while the second path ("test beam") crosses the sample under exam twice while being reflected at a high-quality external mirror. When recombining, the reference and test beams produce an interference pattern, which is recorded with a CCD camera.

Fig. 2. (Color online) Typical maps of the optical path difference measured with the laser interferometric microscope (a) and swelling profile measured with the white-light interferometric microscope (b); the area of interest was implanted with 2 MeV protons at a fluence of $7.63\times10^{16}$ cm$^{-2}$.

Fig. 3. Experimental (dotted) and fitted (line) plots of the optical path difference as a function of implantation fluence of 2 MeV (a) and 3 MeV (b) protons.

Fig. 4. Vacancy density profiles produced in diamond by 2 MeV (continuous line) and 3 MeV (dashed line) protons, as resulting from SRIM Monte Carlo simulation.

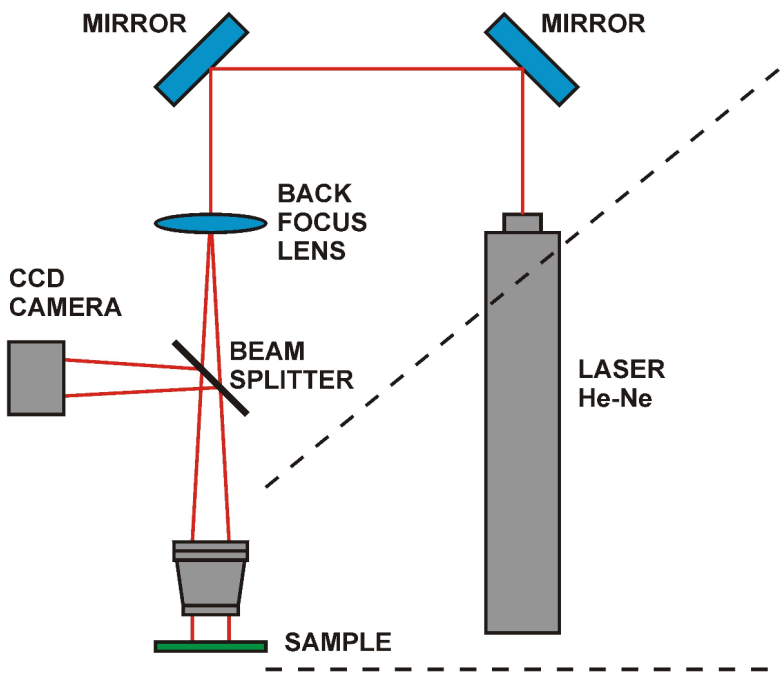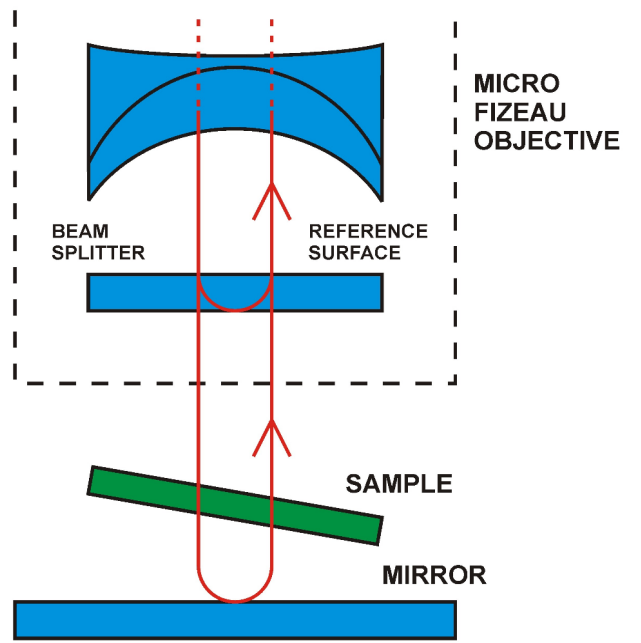

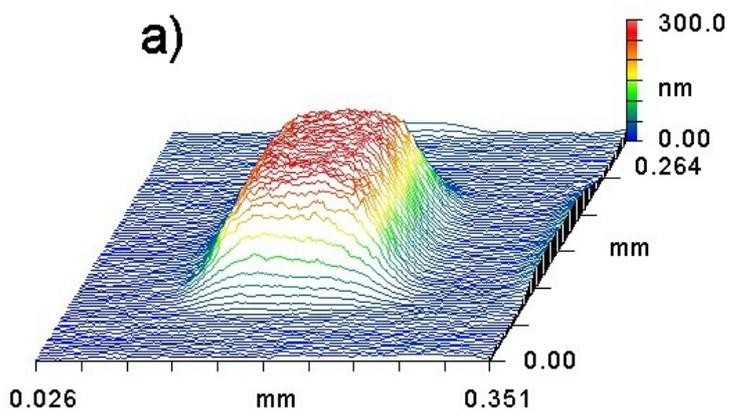 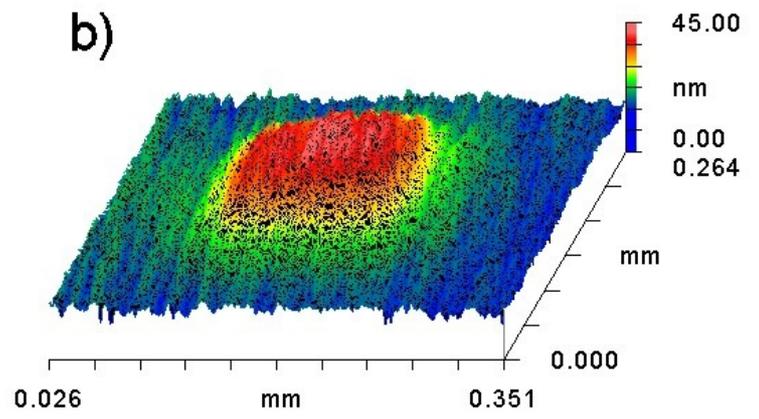

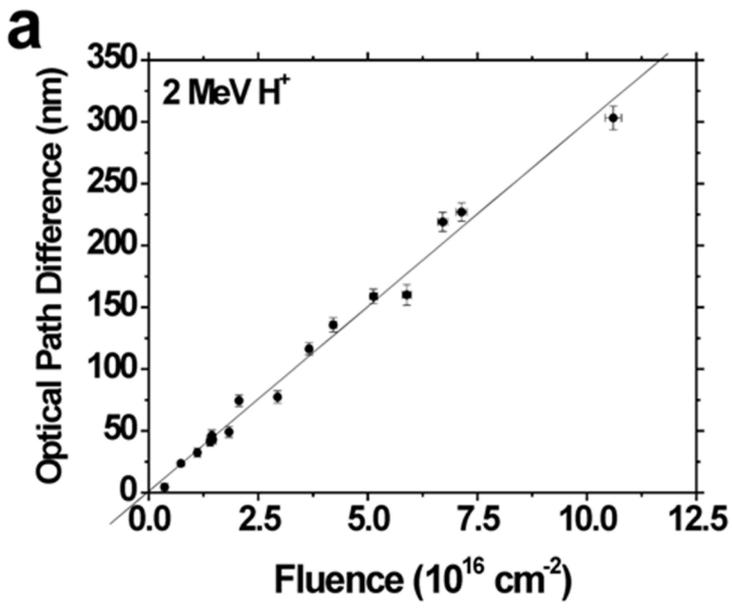 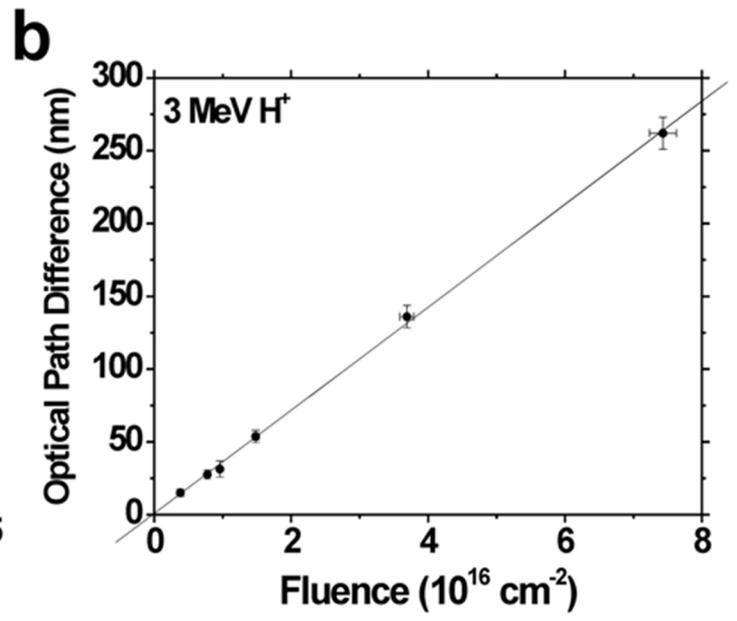

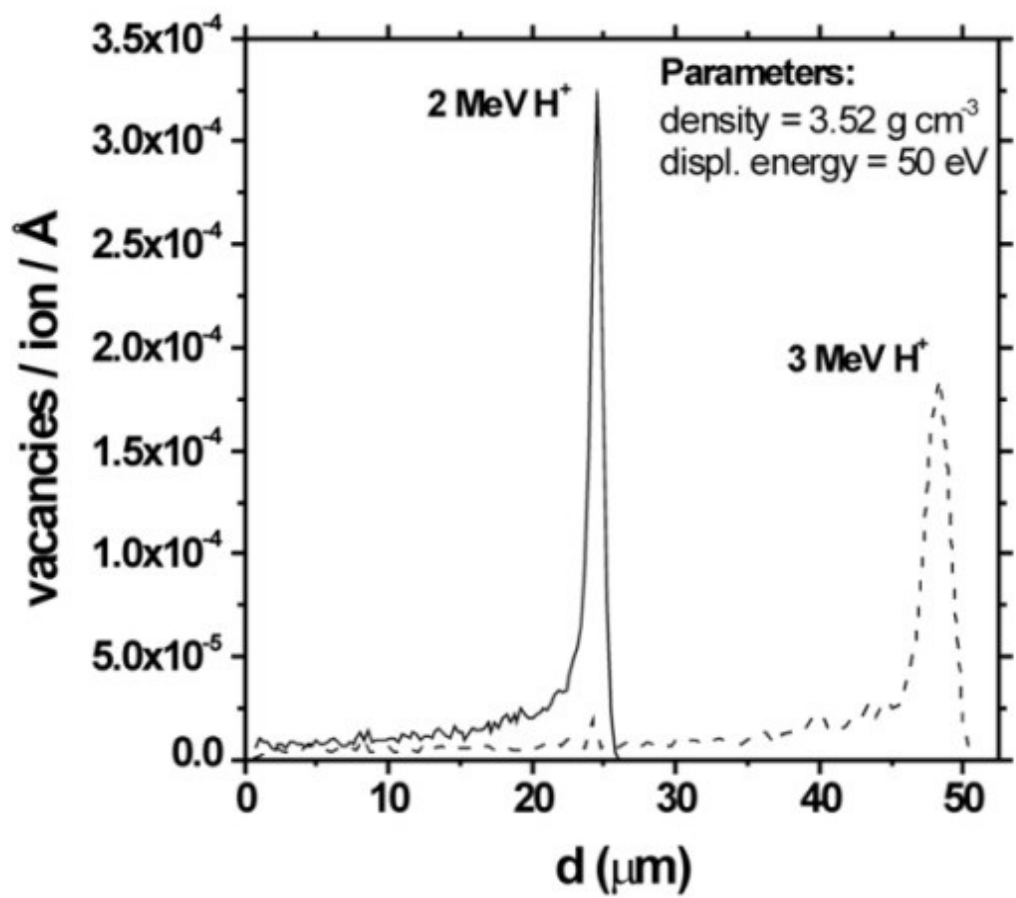